\documentclass[10pt]{article}
\usepackage{amsmath}
\usepackage{amssymb}
\usepackage{bm}
\usepackage{graphicx}
\usepackage[]{natbib}  
\usepackage{color}
\usepackage{comment}

\def\bx{{\bm x}}

\def\memsai{Memorie della Societ\`a Astronomica Italiana}

\def\bx{{\bm x}}

\def\lb{\label}
\def\be{\begin{equation}}
\def\ee{\end{equation}}
\def\bea{\begin{eqnarray}}
\def\eea{\end{eqnarray}}

\def\lb{\label}

\def\be{\begin{equation}}
\def\ee{\end{equation}}
\def\bea{\begin{eqnarray}}
\def\eea{\end{eqnarray}}

\def\bx{{\bm x}}

\def\lb{\label}

\def\be{\begin{equation}}
\def\ee{\end{equation}}
\def\bea{\begin{eqnarray}}
\def\eea{\end{eqnarray}}

\def\aap{Astronomy \& Astrophysics}

\def\bsube{\begin{subequations}}
\def\esube{\end{subequations}}

\begin{document}

\setcounter{figure}{0}
\setcounter{table}{0}
\setcounter{footnote}{0}
\setcounter{equation}{0}

\vspace*{0.5cm}

\noindent {\Large LATEST ADVANCES IN AN ASTROMETRIC MODEL BASED ON THE TIME TRANSFER FUNCTIONS FORMALISM}
\vspace*{0.7cm}

\noindent\hspace*{1.5cm} S. BERTONE$^{1,2}$, A. VECCHIATO$^2$, C. LE PONCIN LAFITTE$^1$, M. CROSTA$^2$, L. BIANCHI$^3$\\

\noindent\hspace*{1.5cm} $^1$ SYRTE - Paris Observatory, UPMC - 77, Av. Denfert Rochereau - 75014 Paris, France\\
\noindent\hspace*{1.5cm} e-mail: stefano.bertone@obspm.fr\\
\noindent\hspace*{1.5cm} $^2$ Astrophysical Observatory of Torino,  Strada Osservatorio 20 - 10025 Pino Torinese, Italy\\
\noindent\hspace*{1.5cm} $^3$ Eurix - Eurixgroup, Via Carcano 26 - 10100 Torino, Italy\\

\vspace*{0.5cm}

\noindent {\large ABSTRACT.} Given the extreme accuracy of modern space astrometry, a precise relativistic modeling of observations is required. 
Moreover, the availability of several models formulated in different and independent ways is a security against the presence of systematic errors in the analysis of future experimental results, like in the case of the Gaia mission.
In this work, we simulate a series of observations using the two models to be used for the data analysis of Gaia, the Gaia RElativistic Model~(GREM) and the Relativistic Astrometric MODel~(RAMOD), and we compare them with the results of our astrometric model based on the Time Transfer Functions.

\vspace*{1cm}

\noindent {\large 1. INTRODUCTION}

\smallskip
A large number of missions planned or proposed for the next years (Gaia, GAME, NEAT) will aim to an increasing astrometric accuracy in order to fulfill their scientific objectives. Aside of the technological capabilities, an accurate description of light propagation in a general relativistic framework will be required. Several independent approaches have been developed in the last years, which are well summarized in~[Vecchiato, 2013]. Then, comparing the outcome of the different models allows for checking their physical consistency.

In particular, the space astrometry mission Gaia~(Bienayme \& Turon, 2002), which has been launched by the European Space Agency (ESA) at the end of 2013, will determine the astrometric parameters for a billion stars with an accuracy of some $\mu as$. The resulting catalog will set the basis for a new celestial reference frame. 
The catalog is built upon a kernel of up to 100 million stars whose positions and motions are reconstructed within a process called "Astrometric Sphere Reconstruction", an extremely difficult task but also a crucial one for the outcome of the mission.
This solution will be performed by the Astrometric Global Iterative Solution (AGIS) software~(Lindegren et al., 2012). At the same time, an independent verification unit for AGIS called Global Sphere Reconstruction (GSR)~(Vecchiato et al., 2012) has been set within the Gaia Data Processing and Analysis Consortium (DPAC). 
Both pipelines are intended to operate on the same real data and the comparison of their results will validate the final astrometric catalog.
In order to keep the two software as separate as possible, two different relativistic modelings of light propagation have been implemented: AGIS relies on GREM~(Klioner, 2003),  while GSR implements RAMOD~(de Felice et al., 2006). 
Moreover, an independent approach to the modeling of the astrometric observables based on the TTF~(Teyssandier et al., 2008) has been recently put in place in the GSR framework. 
An analysis of these three approaches and their application to astrometry has been recently given in~(Bertone et al., 2013). Here, we recall the basis of the TTF approach as well as the early results of the comparison of the three models on simulated Gaia observations.
\vspace*{0.7cm}

\noindent {\large 2. TIME TRANSFER FUNCTIONS IN ASTROMETRY}

\smallskip

The goal of astrometry is to determine the position of celestial bodies from angular measurements. One way to get a relativistic ($i.e.$ covariant) definition of the astrometric observable is by using the tetrad formalism~(Misner et al., 1972), thus giving the direction of observation of an incoming light ray in a particular frame comoving with the observer.

Let us note $E^\mu_{(\alpha)}$ the components of this tetrad, where $(\alpha)$ corresponds to the tetrad index and $\mu$ is a normal tensor index which can be lowered and raised by using the metric. It has been shown in~(Brumberg, 1991) that we can express the direction of the light ray in the tetrad frame as
\begin{equation} \lb{eq:obsTTF}
	n^{(i)}=-\frac{E^0_{(i)}+E^j_{(i)}\hat k_j}{E^0_{(0)}+E^j_{(0)}\hat k_j } = -\frac{E_{(i)}^0+\hat{k}_j E_{(i)}^j}{u^0 \left(1+\hat{k}_j \beta^j \right)}\; ,
\end{equation}
where $\widehat k_i = k_i / k_0$ is the so called light direction triple with $k_\mu = g_{\mu \nu} k^{\nu}$ the covariant components of the tangent vectors to the light ray $k^{\mu} \equiv {dx^\mu}/{d \lambda}$, $u^\alpha$ represents the unit four-velocity of the satellite and $\beta^i \equiv v^i/c$, $v^i$ being the coordinate velocity of the observer.

Let us suppose the existence of a unique light ray connecting the emission event of the signal $x_A = (ct_A, {\bm x}_A)$ and its reception event $x_B = (ct_B, \bx_B)$. It has been shown in~(Bertone \& Le Poncin-Lafitte, 2012) that, under some conditions, $\widehat k_i$ can be expressed as the integral of the metric tensor and its derivatives along the Minkowskian straight line.
Then, a generic tetrad comoving with the chosen observer and computed at the same accuracy can be used in Eq.~\eqref{eq:obsTTF} to compute the direction of light in the observer reference frame.

\vspace*{0.7cm}

\noindent {\large 3. SIMULATED OBSERVATIONS IN THE GAIA CONTEXT}

We implement the model presented in Section~2 in the GSR software and we use it to generate a series of simulated observations.
The result is a "GSR-TTF" code well adapted to support the further development of the GSR code and to investigate the results of both AGIS and GSR.
Let us illustrate how each of Gaia observations is represented in these software.
Each point of the celestial sphere can be fixed in the reference system of the Gaia spacecraft by three direction cosines $n_{(i)}$.
From a geometrical point of view, Gaia will measure the abscissa of such a point, $i.e.$ the angle $\phi$ between the $x$-axis of the spacecraft and the projection of the point in the $x-y$ plane. This angle is related to the director cosines $n_{(i)}$ by the following relations
\be
	\cos\phi = \frac{n_{(1)}}{\sqrt{1- n^2_{(3)}}} \; , \lb{eq:cosphi} 
\ee

The abscissa is generally expressed as function of the astrometric parameters $(\alpha_*,\delta_*, \varpi_*, \mu_{\alpha*}, \mu_{\delta*})$ and of the satellite attitude. Eq.~\eqref{eq:cosphi} also depends on a set of instrument parameters $\{ c_l \}$ to provide a sort of long-term calibration. 
Moreover, when working within the PPN formalism, one should add the parameter $\gamma$ to the unknowns.
As consequence, each of the Gaia observations can be resumed to a non-linear function of these four classes of unknown included in a suitable model of the abscissa $\phi$
\be \lb{eq:obseq}
	\cos \phi \equiv {\cal F} \Big( \alpha_*,\delta_*, \varpi_*, \mu_{\alpha*}, \mu_{\delta*}, a^{(j)}_1,a^{(j)}_2,... ,c_1,c_2,..., \gamma  \Big) \; .
\ee

The GSR software is built so that the director cosines are provided using the RAMOD model (more precisely, the version actually implemented is PPN-RAMOD~(Vecchiato et al., 2003) ). Since the software is built in a modular structure, it is nevertheless possible to use other models to treat light propagation, the aberration corrections, etc.

In particular, we implemented the TTF model presented in Section~2 to compute the director cosines at the accuracy required by the Gaia mission: we modeled the gravitational light deflection within the Solar System by using a PN expansion of the direction triple and the RAMOD tetrad~(Crosta \& Vecchiato, 2010) for the transformation from the Barycentric Celestial Reference System (BCRS) to the Co-Moving Reference System (CoMRS) of Gaia. 
We used the director cosines $n_{(i)}$ so defined to build the abscissae $\phi$, necessary to write the so called known-terms at the left-hand side of Eq.~\eqref{eq:obseq}.

This completes the implementation in GSR of an abscissa $\phi$ based on our model. We shall now compare our results to those of PPN-RAMOD (actually implemented in GSR) and of GREM (the model implemented in AGIS). The analytical equivalence between the models has been shown in~(Bertone et al., 2014): this step will then allow us to validate our implementation and explore the residual differences between the different models.

\begin{figure}[hbt]
\begin{center}
	\includegraphics[width=0.82\textwidth] {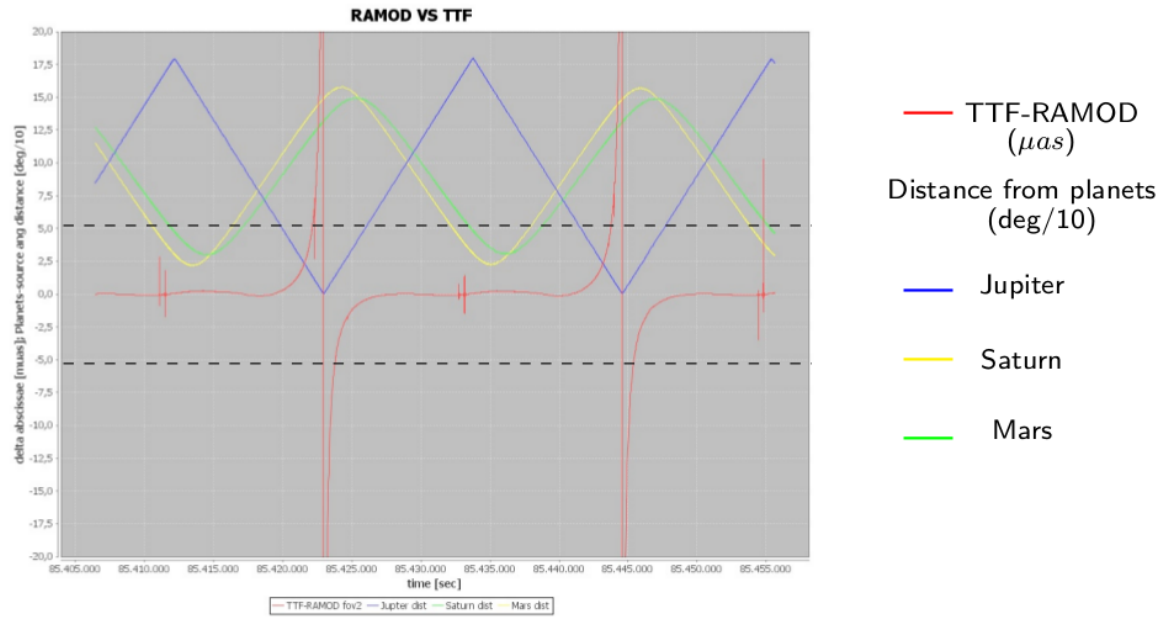} \\
	\includegraphics[width=0.82\textwidth] {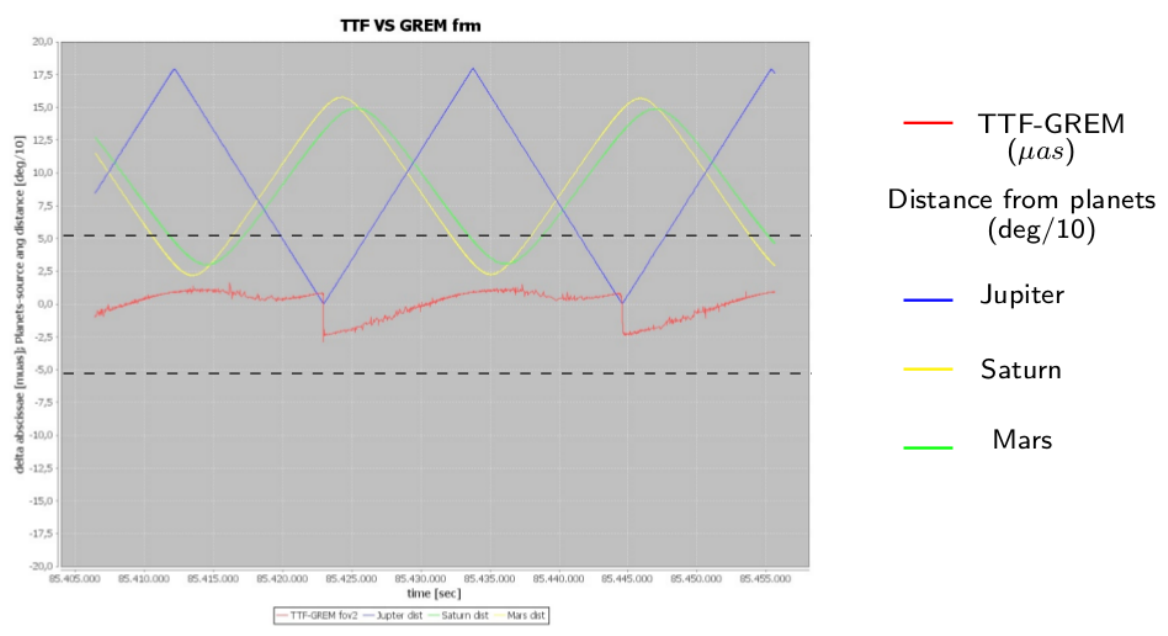} 
\end{center}
\caption{Difference between the abscissae resulting from the TTF and PPN-RAMOD ({\bfseries top}) and GREM ({\bfseries bottom}) models. The numbers on the left axis mark (1) the difference in $\mu as$ between the two models~-~represented by the red plot~-~and (2) the distance in degrees$/10$ between a given planet and the observation. The y-axis is limited to $\pm 20 \; \mu as$ while the black dotted lines indicate a $\pm 5~\mu as$ difference.}
\label{fig:confronto}
\end{figure}

We perform a simulation over one day of observations using the three models to generate the abscissae $\phi$. The results are illustrated in Fig.~\ref{fig:confronto} (produced using the Gaia-tools provided by the Gaia DPAC), where the models are compared one to each other. The numbers on the left axis have a double meaning: they mark (1) the difference in $\mu as$ between the two models~-~represented by the red plot~-~and (2) the distance in degrees$/10$ between a given planet and the observation - the blue, green and yellow plot representing Jupiter, Saturn and Mars, respectively. 
In particular, the periodic oscillation of the distance planet-observation illustrated in the plots is due to the Gaia scanning law~(de Bruijne et al., 2010) setting a rotation period of approximately $6\; h$.
Let us analyze each comparison, noting that we can generally separate the observations in "near" and "far" from the planets with respect to the maximum impact parameter to get $1\; \mu as$ gravitational deflection:
\begin{itemize}
	\item (PPN-)RAMOD vs TTF - we use different models for gravitational light deflection but the same description for the motion of the observer. We get huge differences (up to $500~\mu as$) for the observations near Jupiter. This is expected since PPN-RAMOD is an early study based on a "parametrized" Schwarzschild model of the Solar System, while our TTF model includes the contribution of all major Solar System bodies. On the other hand, the two modelings give equivalent results for most observations "far" from the planets.
	\item TTF vs GREM - both the modelings of gravitational light deflection and the aberration caused by the motion of the observer are different. Nevertheless, as shown in~(Bertone et al., 2014), we would expect not to observe sensible differences in the results. Indeed, the results of the two models are comparable at the $\mu as$ level but the signature of our results far from the Solar System planets suggests substantial differences between the two models or their implementation, while the sudden shift at the maximum approach of the observation field to Jupiter hints for some discrepancy in the treatment of the satellite attitude.
\end{itemize}
 
\vspace*{0.7cm}

\noindent {\large 4. CONCLUSIONS}

This preliminary study shall give a global overview of how the Time Transfer Functions approach can be applied to the complex task of processing the observations of a space astrometry mission.
The periodic signature resulting from the comparison of the abscissae computed using the TTF and GREM approaches, even if limited to the $\mu as$ level, highlights some systematic discrepancy between the two implementations. On the other hand, we observe that the differences between RAMOD and our model are centered around the conjunction with major Solar System bodies. Further investigations will then focus on the implementation of the aberration and the retarded times, which was the same for RAMOD and the TTF but a different one in GREM.
 It also constitutes the basis for the solution of a celestial sphere based on GSR-TTF.

\vspace*{0.7cm}

\textit{Acknowledgements.} S. Bertone thanks UIF/UFI (French-Italian University) for the financial support of this work. S. Bertone and C. Le Poncin-Lafitte are grateful to the financial support of CNRS/GRAM and CNES/Gaia. The work of M. Crosta and A. Vecchiato has been partially funded by ASI under contract to INAF I/058/10/0 (Gaia Mission - The Italian Participation to DPAC).

\vspace*{0.7cm}

\noindent {\large 5. REFERENCES}
%
%
%
%
%
{

\leftskip=5mm
\parindent=-5mm

\smallskip

Bertone, S. \& Le Poncin-Lafitte, C. 
, \memsai, 2012, 83, 1020

Bertone, S.; Le Poncin-Lafitte, C.; Crosta, M.; Vecchiato, A.; Minazzoli, O. \& Angonin, 
, SF2A-2013: Proceedings of the Annual meeting of the SF2A, 2013, 155-159

Bertone, S.; Minazzoli, O.; Crosta, M.; Le Poncin-Lafitte, C.; Vecchiato, A. \& Angonin, M.-C. 
, Classical and Quantum Gravity, 2014, 31, 015021

Bienayme, O. \& Turon, C. 
, EAS Publications Series, 2002, 2

Brumberg, V. A. 
, Bristol, England and New York, Adam Hilger, 1991, 271 p., 1991

Crosta, M. \& Vecchiato, A. 
, Astronomy and Astrophysics, 2010, 509, A37

de Bruijne, J.; Siddiqui, H.; Lammers, U.; Hoar, J.; O'Mullane, W. \& Prusti, T. Klioner, S. A.
, IAU Symposium, 2010, 261, 331-333

de Felice, F.; Vecchiato, A.; Crosta, M. T.; Bucciarelli, B. \& Lattanzi, M. G. 
, Astrophysical Journal, 2006, 653, 1552-1565

Klioner, S. A. 
, Space Astronomical Journal, 2003, 125, 1580-1597

Lindegren, L.; Lammers, U.; Hobbs, D.; O'Mullane, W.; Bastian, U. \& Hern\`andez, J. 
, \aap, 2012, 538, A78

Misner, C.; Thorne, K. \& Wheeler, J. Gravitation San Francisco: W. H. Freeman, 1973

Teyssandier, P. \& Le Poncin-Lafitte, C. 
, Classical and Quantum Gravity, 2008, 25, 145020

Vecchiato, A.; Lattanzi, M. G.; Bucciarelli, B.; Crosta, M.; de Felice, F. \& Gai, M. 
, \aap, 2003, 399, 337-342

Vecchiato, A.; Abbas, U.; Bandieramonte, M.; Becciani, U.; Bianchi, L.; Bucciarelli, B.; Busonero, D.; Lattanzi, M. G. \& Messineo, R. 
, Society of Photo-Optical Instrumentation Engineers (SPIE) Conference Series, 2012, 8451

Vecchiato, A.; Gai, M.; Lattanzi, M. G.; Crosta, M.; Becciani, U. \& Bertone, S. 
, ArXiv e-prints, 2013

}

\end{document}